\documentclass[pra,10pt,amsmath,twocolumn]{revtex4-1}
\usepackage{amsmath}
\usepackage{amsfonts}
\usepackage{graphicx}

\newcommand{\op}[1]{\operatorname{#1}}

\newcommand{\parti}[2]{\frac{\partial #1}{\partial #2}}
\newcommand{\partit}[2]{\frac{\partial^2 #1}{\partial #2^2}}
\newcommand{\diff}[2]{\frac{d #1}{d #2}}

\newcommand{\intall}{\int_{-\infty}^{\infty}}
\newcommand{\ket}[1]{|#1\rangle}

\newcommand{\bra}[1]{\langle#1|}

\newcommand{\avg}[1]{\langle#1\rangle}

\newcommand{\sech}{\operatorname{sech}}

\newcommand{\abs}[1]{\left|#1\right|}
\newcommand{\bk}[1]{\left(#1\right)}
\newcommand{\Bk}[1]{\left[#1\right]}
\newcommand{\BK}[1]{\left\{#1\right\}}

\newcommand{\trace}{\operatorname{tr}}

\begin{document}
\title{Quantum transition-edge detectors}

\author{Mankei Tsang}

\email{eletmk@nus.edu.sg}
\affiliation{Department of Electrical and Computer Engineering,
  National University of Singapore, 4 Engineering Drive 3, Singapore
  117583}

\affiliation{Department of Physics, National University of Singapore,
  2 Science Drive 3, Singapore 117551}









\date{\today}

\begin{abstract}
  Small perturbations to systems near critical points of quantum phase
  transitions can induce drastic changes in the system properties.
  Here I show that this sensitivity can be exploited for weak-signal
  detection applications. This is done by relating a widely studied
  signature of quantum chaos and quantum phase transitions known as
  the Loschmidt echo to the minimum error probability for a quantum
  detector and noting that the echo, and therefore the error, can be
  significantly reduced near a critical point. Three examples, namely,
  the quantum Ising model, the optical parametric oscillator model,
  and the Dicke model, are presented to illustrate the concept. For
  the latter two examples, the detectable perturbation can exhibit a
  Heisenberg scaling with respect to the number of detectors, even
  though the detectors are not entangled and no special quantum state
  preparation is specified.
\end{abstract}

\maketitle
Phase transitions are characterized by macroscopic changes to a system
due to slight variations in the system parameters. Sensing is a
natural application of this sensitivity. For example, superconducting
transition-edge sensors exploit the highly temperature-sensitive
resistance of a superconductor near the critical temperature to
measure energy deposition and can detect photons with record
efficiency \cite{irwin}. Such devices rely on classical phase
transitions, which are sensitive to thermodynamic variables only. This
limitation rules out the use of classical phase transitions for many
sensing applications, such as optical interferometry, force sensing,
accelerometry, gyroscopy, and magnetometry, where the signals of
interest can barely perturb the thermodynamic variables. Here I
propose the concept of quantum transition-edge sensors, which exploit
the sensitivity of quantum phase transitions to Hamiltonian parameters
\cite{sachdev} and should thus be useful for a much wider range of
quantum sensing and system identification applications. On a
fundamental level, the feasibility of the proposal is demonstrated by
relating a well known measure of quantum chaos and quantum phase
transitions known as the Loschmidt echo
\cite{peres1984,*gorin,*goussev} to the theoretical minimum error
probability for a quantum detector \cite{helstrom}.  A small echo then
directly implies that an optimal measurement of the system can
accurately detect the perturbation.  Three examples, namely, the
quantum Ising model \cite{sachdev}, the optical parametric oscillator
(OPO) model \cite{walls_milburn}, and the Dicke model
\cite{dicke,emary_pre}, are presented to illustrate the concept.


Suppose that the initial quantum state is $\ket{\psi}$ and the
Hamiltonian is $H_0$. After time $t$, a Hamiltonian $-H_1$ is applied
to reverse the evolution.  The final state given by
\begin{align}
\ket{\psi'} &= U_1^\dagger U_0\ket{\psi},
&
U_{m} &\equiv \mathcal T\exp\Bk{-i\int_0^t d\tau H_m(\tau)},
\end{align}
where $m\in\{0,1\}$, should be different from $\ket{\psi}$ if $H_0\neq
H_1$. A measure of the difference is the overlap between the initial
and final states called the Loschmidt echo
\cite{peres1984,*gorin,*goussev}:
\begin{align}
F &\equiv \abs{\avg{\psi'|\psi}}^2 = 
\abs{\bra{\psi}U_0^\dagger U_1\ket{\psi}}^2.
\label{echo}
\end{align}
The echo is a measure of how accurately the dynamics of a quantum
system can be reversed by an imperfect time-reversal operation. An
enhanced decay of the echo with respect to a given difference between
$H_1$ and $H_0$ can be used as a signature of quantum chaos
\cite{peres1984,*gorin,*goussev} and criticality \cite{quan}, when
time reversibility is compromised.

Let us now consider a different scenario more conducive to weak-signal
detection applications: the initial state is again $\ket{\psi}$, but
imagine that there are two possibilities for the Hamiltonian, namely,
$H_0$ or $H_1$. For detection problems, assume that $H_0$ is the
unperturbed Hamiltonian and $H_1$ is the perturbed one.  The final
state is either $\ket{\psi_0} = U_0 \ket{\psi}$ or $\ket{\psi_1} = U_1
\ket{\psi}$.  A measurement is then performed, with outcome $y$, to
detect the presence of the perturbation. The probability of $y$ given
either hypothesis is $P(y|\mathcal H_m) = \trace\Bk{E(y)\rho_m}$,
where $E(y)$ is a positive operator-valued measure (POVM), the most
general way of specifying the statistics of a quantum measurement
\cite{nielsen}, $\rho_m = \ket{\psi_m}\bra{\psi_m}$ is a density
operator, and $\mathcal H_m$ denotes the hypothesis. In the context of
quantum information theory, this is known as the unitary
discrimination problem
\cite{ou,*paris1997,childs,*acin2001,*ajv,*dariano,tsang_nair,tsang_open}.
A general decision strategy entails separating the space of $y$ into
two regions $\Upsilon_0$ and $\Upsilon_1$; if $y$ is in $\Upsilon_0$
one decides that $\mathcal H_0$ is true and vice versa. Let the error
probabilities be
$P_{10} \equiv \int_{\Upsilon_1} dy P(y|\mathcal H_0)$ and
$P_{01} \equiv \int_{\Upsilon_0} dy P(y|\mathcal H_1)$.
Given prior probabilites $P_0$ and $P_1$, the average error
probability is $P_e \equiv P_{10}P_0 + P_{01}P_1$.  A seminal result
by Helstrom \cite{helstrom} states that the minimum $P_e$ for any POVM
is
\begin{align}
\min_{E(y)} P_e &= \frac{1}{2}\bk{1-\sqrt{1-4P_0P_1F}},
\label{helstrom}
\end{align}
where
\begin{align}
F &\equiv \abs{\avg{\psi_0|\psi_1}}^2 =
\abs{\bra{\psi}U_0^\dagger U_1\ket{\psi}}^2
\end{align}
is called the fidelity, which is exactly the same as the Loschmidt
echo given by Eq.~(\ref{echo}). $\min_{E(y)} P_e$ decreases
monotonically with decreasing $F$. It follows that, whenever $F \ll 1$
such that the $\min_{E(y)} P_e \approx P_0P_1 F \ll 1$, there exists a
measurement that enables one to distinguish the two hypotheses and
detect the perturbation accurately. Conversely, if $F$ is high such
that $\min_{E(y)} P_e$ is high, no measurement can accurately tell the
two hypotheses apart.

Let $H_0 = H(x)$ and $H_1 = H(x+\delta)$, where $x$ and $\delta$ are
continuous parameters. $G_{jk} \equiv -2\partial^2 F/\partial
\delta_{j}\partial \delta_{k}\big|_{\delta=0}$ is called the quantum
Fisher information, the inverse of which can be used to lower-bound
the mean-square error in estimating $x$ via the quantum Cram\'er-Rao
bound (QCRB) \cite{helstrom,hayashi,*paris,twc}. $G$, like $F$, has
been used to study quantum criticality \cite{zanardi_prl}, but $G$ is
a less conclusive measure from the perspective of quantum metrology,
as the attainment of the QCRB may require repeated adaptive
measurements \cite{fujiwara2006} that can negate the advantage of
having a high $G$ \cite{qzzb,*gm_useless}, unlike the one-shot
attainability of the Helstrom bound. Although $G$ also determines the
behavior of $F$ near $F = 1$ via the approximation $F \approx
1-\delta^\top G\delta/4$, in the following I shall focus on the more
useful $F\ll 1$ regime, where accurate detection is possible and $G$
has little relevance.

The connection between fidelity measures for quantum phase transitions
and quantum metrology was also pointed out by
Refs.~\cite{zanardi2008,*invernizzi}, while the use of thermal states
near a quantum critical point of a Dicke-Ising model for metrology was
proposed by Ref.~\cite{gammelmark}, but they all focused on states at
thermal equilibrium and not the dynamics. For sensing, time is often a
limited resource due to a finite signal duration or deteriorating
experimental conditions, so the dynamical response of a sensor, the
main focus here, is more important and relevant than the equilibrium
properties studied in previous work. On a foundational level, time is
of course such a fundamental physical quantity that makes the
finite-time quantum-information-theoretic measures interesting in
their own right.  Another relevant prior work is
Ref.~\cite{cucchietti}, which proposed a Loschmidt echo experiment
with a Bose-Hubbard system for sensing applications but have not
studied the fundamental sensitivity enabled by the system.

Before studying specific examples, it is helpful to first recall a
standard solution for $F$ in quantum detection theory \cite{helstrom}
for comparison. Suppose that $\ket{\psi} = \ket{\phi}^{\otimes N}$,
$H_0 = q x_0$, $H_1 = q(x_0+\delta)$, $x_0$ and $\delta$ are scalars,
and $\ket{\phi}$ has a Gaussian distribution with respect to
eigenstates of $q$. Then
\begin{align}
F &= \exp\bk{-N \Delta q^2 \delta^2 t^2},
\label{F}
\end{align}
where $\Delta q^2$ is the variance of $q$ for $\ket{\phi}$.  In
detection applications, one is usually interested in the error
exponent $-\ln P_e$ as measure of detection performance and desire
$-\ln P_e \gg 1$. In this low-error regime, the optimal error exponent
is
\begin{align}
\max_{E(y)} \bk{-\ln P_e} &\approx -\ln P_0-\ln P_1  -\ln F,
\label{optimal_exp}
\end{align}
which differs from the fidelity exponent $-\ln F$ by just a constant
factor. I shall hereafter focus on $-\ln F$ as a figure of
merit. Given Eq.~(\ref{F}), the fidelity exponent is
\begin{align}
-\ln F &= N \Delta q^2 \delta^2 t^2.
\label{shot}
\end{align}
Another useful performance measure is called the detectable
perturbation $\delta'$ \cite{ou,*paris1997,tsang_nair,tsang_open},
which is the magnitude of $\delta$ that leads to an acceptable error
probability $P_e'$.  Defining $F'$ as the fidelity that leads to $P_e
= P_e'$ via Eq.~(\ref{helstrom}), one obtains
\begin{align}
\delta' &= \frac{\sqrt{-\ln F'}}{\sqrt{N}\Delta q t}.
\label{shot_delta}
\end{align}
$\delta'$ quantifies the sensitivity of a detector with respect to
resources $N$ and $t$.  I define the scalings of Eqs.~(\ref{shot}) and
(\ref{shot_delta}) with respect to $N$, $\delta$, and $t$ as the
standard scalings.

As the first example, consider the quantum Ising model \cite{sachdev}:
\begin{align}
H_m &= -J\sum_{j=1}^N \bk{\sigma_j^z\sigma_{j+1}^z +g_m \sigma_j^x},
\end{align}
where $\sigma_j^x$ and $\sigma_j^z$ are Pauli spin operators, $J$ is
the spin interaction strength, $g$ is the transverse magnetic field
normalized with respect to $J$, and the periodic boundary condition is
assumed. Let $\delta \equiv g_1 - g_0$ be the
perturbation. Conventional quantum metrology protocols prepare
$\ket{\psi}$ in a special state and then apply a simple Hamiltonian
$\propto \sum_j \sigma_j^x$ \cite{glm_science}. Here I simply assume
$\ket{\psi}$ to be the ground state of $H_0$; the additional terms in
the Hamiltonian may be regarded as coherent quantum control
\cite{lloyd2000,*mabuchi,*wiseman_milburn} in place of state
preparation. An analytic solution for $F$ is \cite{quan}
\begin{align}
F &= \prod_{k=1}^{N/2}\BK{1-\sin^2[\epsilon_1(k) t]\sin^2\Bk{\theta_1(k)-\theta_0(k)}},
\label{exact_F}
\\
\epsilon_1(k) &\equiv 2J\sqrt{1+g_1^2 - 2g_1\cos \phi(k)},
\\
\theta_m(k) &\equiv \op{arctan} \frac{\sin\phi(k)}{g_m-\cos\phi(k)},
\quad
\phi(k) \equiv \frac{2\pi k}{N}.
\end{align}
Heuristic and numerical analyses in Ref.~\cite{quan} suggest that the
decay of $F$ with respect to $\delta$ is enhanced near the critical
point $g = 1$. Using a similar analysis and relating $F$ to the
product yield in a chemical reaction, Ref.~\cite{cai2012} also
suggests that the criticality may be useful for avian magnetometry
\cite{lambert}. Here I study $F$ more carefully in the thermodynamic
limit ($N\to\infty$), similar to the calculation done for a different
purpose in Ref.~\cite{silva}. Assume that each Bloch mode contributes
little to the decay of $F$, and
\begin{align}
\epsilon_1 t &\ll 1, 
&
\sin^2(\epsilon_1 t) &\approx \epsilon_1^2 t^2,
\label{short_time}
\end{align}
which can be justified in the $N\to\infty$ limit, as will be shown
later.  For a small enough $\delta$, $\theta_1-\theta_0$ can be
approximated in the first order according to $\op{arctan}(1/x_0)
\approx \op{arctan} (1/x_1) -(x_0-x_1)/(1+x_1^2)$. Assuming further
that $\delta$ is small enough such that $|\theta_1-\theta_0| \ll 1$
and $\ln[1-\sin^2(\epsilon_1t)\sin^2(\theta_1-\theta_0)]\approx
-\epsilon_1^2 t^2(\theta_1-\theta_0)^2$, one obtains
\begin{align}
-\ln F &\approx 4J^2\delta^2t^2 \sum_{k=1}^{N/2} 
\frac{ \sin^2 \phi}{1+g_1^2-2g_1\cos\phi}.
\end{align}
In the $N\to\infty$ limit, the discrete sum over Bloch modes can be
replaced with an integral with respect to $\phi$:
\begin{align}
-\ln F &\approx 4J^2\delta^2t^2  \frac{N}{2\pi}\int_0^\pi d\phi
\frac{ \sin^2 \phi}{1+g_1^2-2g_1\cos\phi}
\\
&= \Big\{\begin{array}{cc}
NJ^2\delta^2t^2/g_1^2, & g_1 > 1,\\
NJ^2\delta^2t^2, & g_1 \le 1.
\end{array}
\label{ising_expo}
\end{align}
With this result, Eq.~(\ref{short_time}) can now be justified by
noting that any value of $F > 0$ can be reached by setting
the time as
\begin{align}
t &= \frac{\sqrt{-\ln F}}{\sqrt{N} J \delta}\times
\Big\{\begin{array}{cc} g_1, & g_1 > 1,\\
1, & g_1 \le 1,
\end{array}
\end{align}
which scales with $1/\sqrt{N}$. Thus, given $F$, $J$, and $\delta$,
one can always increase $N$ and find a time that satisfies
Eqs.~(\ref{short_time}).

The nonanalyticity of $F$ at $g_1 = 1$ indicates a quantum phase
transition. Unfortunately for metrology, Eq.~(\ref{ising_expo}) has
the same scalings with repsect to $N$, $\delta$, and $t$ as the
standard limit given by Eq.~(\ref{shot}).  This result means that the
quantum Ising model in the thermodynamic limit does not provide any
enhancement beyond the standard limit for magnetometry.

The next two examples, both of which involve bosonic rather than
fermionic excitations, turn out to be far more promising. Consider
first the model for a degenerate OPO under threshold
\cite{walls_milburn}:
\begin{align}
H_m &= \omega_m a^\dagger a + i\lambda_m\bk{a^{\dagger2} - a^2},
\label{opo}
\end{align}
where $a$ and $a^\dagger$ are bosonic annihilation and creation
operators, $\omega_m$ is the frequency detuning, which can be
perturbed by the motion of the cavity mirrors or phase shifts inside
the optical cavity, and $\lambda_m$ is the parametric pump strength,
assumed to be a $c$-number. This assumption, common in quantum optics,
is valid when the pump is strong and undepleted. Define the
criticality parameter as $g_m = 2\lambda_m/\omega_m$, with $\lambda_m$
assumed to be real. Assume that the system is biased in such a way
that $g_0 < 1$, for which the system is below threshold, and the
perturbation $\delta$ would cause $g_1 = g_0+\delta > 1$ and thus
instability. For example, a small change $\Delta\omega \equiv
\omega_1-\omega_0$ in the detuning, with $\lambda_0 = \lambda_1$ held
fixed, induces a perturbation $\delta \approx -2\lambda_0
\Delta\omega/\omega_0^2$. $H_0$ can be diagonalized using the
Bogoliubov transformation:
\begin{align}
b_0 &= \mu_0 a + i\nu_0 a^\dagger,
\\
\nu_0 &= \frac{1}{\sqrt{2}}\sqrt{(1-g_0^2)^{-1/2}-1},
\quad
\mu_0 = \sqrt{1+\nu_0^2},
\\
H_0 &= \omega' b_0^\dagger b_0 + E_0,
\quad
\omega' \equiv \omega_0\sqrt{1-g_0^2},
\end{align}
where the ground-state energy $E_0$ is irrelevant to subsequent
calculations.

If $\ket{\psi}$ is the ground state of $H_0$ and $H_1$ is applied, the
system becomes unstable, initiating a transition to the oscillation
phase \cite{walls_milburn}.  Until the pump is depleted significantly,
there is still a period of time over which Eq.~(\ref{opo}) is
accurate.  The Hamiltonian can then be expressed by
\begin{align}
b_1 &= \mu_1 a + i \nu_1 a^\dagger = \mu' b_0 + i\nu' b_0^\dagger,
\\
\nu_1 &= \frac{1}{\sqrt{2}}\sqrt{(1-g_1^{-2})^{-1/2}-1},
\quad
\mu_1 = \sqrt{1+\nu_1^2},
\\
\mu' &= \mu_1\mu_0-\nu_1\nu_0,
\quad
\nu' = \nu_1\mu_0-\mu_1\nu_0,
\\
H_1 &= i\lambda'\bk{b_1^{\dagger 2}-b_1^2} + E_1,
\quad
\lambda' \equiv \lambda_1\sqrt{1-g_1^{-2}},
\end{align}
where $E_1$ is another unimportant scalar. $\bra{\psi}U_0^\dagger
U_1\ket{\psi} = \bra{\psi}\exp(-iH_1t)\ket{\psi}$ can be computed by
writing $H_1$ in terms of $b_0$ and invoking the $SU(1,1)$
disentangling theorem \cite{santiago}. The result is
\begin{align}
F &= \Bk{1+\bk{1+2\nu'^2}^2 \sinh^2(2\lambda' t)}^{-1/2},
\end{align}
which decreases with increasing $\lambda't$ and $\nu'$.  If $g_1$ is
just above the critical point with $g_1 = 1+\delta_1/2$ and
$0<\delta_1 \ll 1$, $\lambda' \approx \lambda_1\sqrt{\delta_1}$.  For
$\nu'$, the worst case is when $\delta_1 = \delta$ and $g_0 =
1-\delta/2$ such that $\nu' \approx 0$, which leads to
\begin{align}
F &\approx \sech(2\lambda_1\sqrt{\delta}t).
\label{worst_F}
\end{align}
In the limit of $2\lambda_1\sqrt{\delta}t \gg 1$,
\begin{align}
-\ln F &\approx 2\lambda_1\sqrt{\delta} t - \ln 2,
\end{align}
which scales with the much larger $\sqrt{\delta}$ rather than the
$\delta^2$ standard scaling in Eq.~(\ref{shot}) (since $\delta \ll 1$),
although the time dependence here is linear rather than quadratic. The
detectable perturbation given by
\begin{align}
\delta' &\approx \frac{[-\ln(F'/2)]^2}{4\lambda_1^2t^2}
\end{align}
decreases with time as $1/t^2$, which is quicker than the $1/t$
standard scaling in Eq.~(\ref{shot_delta}). These results confirm the
intuition that quantum criticality can enhance the sensitivity of a
detector to weak perturbations.

A near-optimal measurement analogous to Kennedy's receiver for
coherent-state discrimination \cite{helstrom,kennedy} can be realized
by counting photons in the $b_0$ mode.  Let $E(n) = \ket{n}\bra{n}$,
where $\ket{n}$ is an eigenstate of $b_0^\dagger b_0$ with
$b_0^\dagger b_0\ket{n} = n\ket{n}$. Under $\mathcal H_0$, the count
is always zero, and under $\mathcal H_1$, the probability of counting
$n$ photons is
\begin{align}
P(n|\mathcal H_1) = |\bra{n}\exp(-iH_1t)\ket{0}|^2.
\end{align}
If one decides on $\mathcal H_0$ when $n = 0$ and on $\mathcal H_1$
when $n > 0$,
\begin{align}
P_{10} &= 0,
\quad
P_{01} = |\bra{0}\exp(-iH_1t)\ket{0}|^2 = F,
\\
-\ln P_e &= -\ln P_1 - \ln F.
\end{align}
This error exponent is smaller than the optimal value in
Eq.~(\ref{optimal_exp}) by just a constant $-\ln P_0$.

The calculations so far are accurate only when the undepleted pump
approximation is valid, and for long enough time the final state under
$\mathcal H_1$ is expected to stabilize, leading to a saturating
$F$. This is not a problem, however, as long as the desirable $P_e$ is
reached before the saturation occurs; the saturation can be delayed by
reducing the parametric coupling strength and increasing the pump
power.

Instead of one OPO mode, consider $N$ such modes, and assume that each
mode contributes little to the decay of $F$, such that
$\sech(2\lambda_1\sqrt{\delta}t)\approx 1-2\lambda_1^2\delta t^2$.
The collective fidelity and detectable perturbation become
\begin{align}
F &\approx (1-2\lambda_1^2\delta t^2)^N \approx
\exp(-2N\lambda_1^2\delta t^2),
\\
-\ln F &\approx 2N\lambda_1^2\delta t^2,
\quad
\delta' \approx \frac{-\ln F'}{2N\lambda_1^2 t^2}.
\end{align}
The fidelity exponent now scales with $t^2$. It is even more
intriguing to see the $1/N$ ``Heisenberg'' scaling for $\delta'$
enabled by the quantum criticality, even though the modes are not
entangled. Using a large $N$ can also alleviate the saturation
problem, as one can reduce the detection time and avoid saturation by
increasing $N$.

We can consider an even more practical measurement model by
introducing traveling fields that couple to the OPO and continuous
measurements, such as heterodyne detection \cite{gardiner_zoller}.
The constant coupling, however, is expected to damp the instability
and cause suboptimal behavior. The Supplementary Material \cite{sup}
contains a detailed calculation of the classical Fisher information
$\mathcal G(\omega_m)$ for the estimation of the resonance frequency
$\omega_m$ for such a model with $N=1$ and $\lambda$ held fixed. The
classical Fisher information is an acceptable metrological measure
here because the mean-square error can approach $\mathcal G^{-1}$ in a
large-deviation limit using maximum-likelihood estimation \cite{sato},
which is easy to perform numerically in practice
\cite{shumway_stoffer,*ang}. The calculation shows that, despite the
damping and the suboptimal heterodyne measurement, $\mathcal G$ can be
enhanced by orders of magnitude as $g$ approaches the OPO
threshold. At the threshold, $4\omega_m^2t/\gamma^3 >\mathcal G >
1.532\omega_m^2 t/\gamma^3$, where $\gamma$ is the coupling rate and
$0<\gamma < 4\lambda$ is assumed. As expected, $\gamma$ limits the
Fisher information, but it also means that a reduction of $\gamma$ can
enhance the information significantly. With the advent of
ultrahigh-quality optical resonators
\cite{vahala,*kippenberg,*safavi2012} and their experimentally
demonstrated parametric instabilities \cite{kippenberg2004,*rokhsari},
this enhancement of Fisher information implies that the concept of
transition-edge sensors is immediately relevant to current technology,
even if the quantum-optimal scalings are less trivial to attain.

As the final example, consider the Dicke model
\cite{dicke,emary_pre}. An experimental demonstration of the Dicke
quantum phase transition was recently reported in
Ref.~\cite{baumann}. In the normal phase, the Hamiltonian can be
approximated as \cite{emary_pre}
\begin{align}
  H_m &\approx \omega_m a^\dagger a + \omega_m b^\dagger b +
  i\lambda_m(a^\dagger + a) (b^\dagger + b),
\end{align}
where $a$ and $b$ are annihilation operators of two bosonic modes and
their frequencies are assumed to be the same for simplicity.  The
criticality parameter is $g_m = 2\lambda_m/\omega_m$, and the critical
point is $g_m = 1$.  Assume again that $g_0 = 1-\delta/2 < 1$, $g_1 =
1+\delta/2 > 1$, $\ket{\psi}$ is the ground state of $H_0$, and the
normal-phase approximation of the Hamiltonian is accurate for the time
considered. $H_0$ can be diagonalized in the form of $\epsilon_+
c_+^\dagger c_+ + \epsilon_- c_-^\dagger c_-$, where $c_\pm$ are the
normal-mode bosonic operators, whereas $H_1$ can be expressed in the
form of $\epsilon_{1+}c_{1+}^\dagger c_{1+} + (\lambda'c_{1-}^{\dagger
  2}+\lambda'^*c_{1-}^2)$, with $c_{1+}$ a function of $c_+$ and
$c_{1-}$ a function of $c_-$, indicating that the $c_-$ mode becomes
unstable. Using the same techniques mentioned in the previous example,
it can be shown that the resulting fidelity is
\begin{align}
F &\approx F_+\sech(\omega_1\sqrt{\delta}t),
\end{align}
where $0 < F_+ \le 1$ is a factor that oscillates with time due to the
$c_+$ mode \cite{*[{}] [{ computed $F$ for the Dicke model
    with $g_1 < 1$, demonstrating the oscillating behavior of $F$ for
    stable modes, but they did not consider the $g_1 > 1$ case.}]
  paraan}. Similar to the OPO example, $-\ln F$ scales with
$\sqrt{\delta} t$, rather than the standard scaling $\delta^2 t^2$ in
Eq.~(\ref{shot}).  A similar behavior is expected if $H_0$ is the
superradiant-phase approximation and $H_1$ initiates a transition to
the normal phase. These results suggest that bosonic phase transitions
can offer significant accuracy improvements for weak-signal detection.

I have outlined the fundamental principles of quantum transition-edge
detectors, but many open questions remain. Practical implementations
and the effects of excess noise and decoherence in particular deserve
further study, and may be analyzed using the methods in
Refs.~\cite{fujiwara,*escher,*escher_bjp,*escher_prl,*kolodynski,*demkowicz,*kolodynski2013,*knysh,tsang_open}. Quantum
control methodologies \cite{lloyd2000,*mabuchi,*wiseman_milburn} may
be useful for finding the best Hamiltonians and measurements that
optimize the sensitivity in practice.  Sensitivity of quantum systems
to multiparameter, time-dependent, or stochastic perturbations
\cite{twc,tsang_nair,tsang_open} is another interesting problem and
may be enhanced by non-equilibrium quantum phase transitions
\cite{bastidas,*bastidas_pra,*engelhardt}.


Helpful discussions with N.~Lambert, F.~Nori, S.~K.~Ozdemir,
V.~M.~Bastidas, and X.~Wang are gratefully acknowledged. This work is
supported by the Singapore National Research Foundation under NRF
Grant No. NRF-NRFF2011-07.

\bibliography{research}

\appendix

\section{Supplemental Material}
Consider a degenerate optical parametric oscillator (OPO)
\cite{gardiner_zoller}.  The equation of motion for the optical-mode
analytic signal is
\begin{align}
\diff{a(t)}{t} &= -\frac{\gamma}{2}a(t) -i\omega_m a(t) + 2\lambda a^*(t) + \sqrt{\gamma} A(t),
\end{align}
where $\gamma$ is the coupling rate, $\omega_m$ is the resonance
frequency, $\lambda$ is the pump coefficient, and $A(t)$ is the input
field.  The output field is given by
\begin{align}
A_{\rm out}(t) &= \sqrt{\gamma} a(t) - A(t) + A'(t),
\end{align}
where $A'$ is an excess noise. Suppose that $A_{\rm out}$ is
measured by continuous heterodyne detection, and $A$ and $A'$ are
white phase-insensitive noises with noise powers $S_{\rm in}$ and
$S'$. After some lengthy but standard calculations, the
output power spectral density is given by
\begin{align}
S(\omega) &= [1+2V(\omega)]S_{\rm in} + S',
\end{align}
where $V(\omega)$ is the idler gain. In terms of normalized frequency
and parameters,
\begin{align}
V(\Omega) &= \frac{\Gamma^2}
{\Bk{\Omega^2 -\bk{g^{-2}-1-\Gamma^2/4}}^2+(g^{-2}-1)\Gamma^2},
\\
\Omega &\equiv \frac{\omega}{2|\lambda|},
\quad
g \equiv \frac{2|\lambda|}{\omega_m},
\quad
\Gamma \equiv \frac{\gamma}{2|\lambda|}.
\end{align}
To compute the Fisher information for estimating $\omega_m$
from $A_{\rm out}$, we start with the Bhattacharyya distance \cite{vantrees3}:
\begin{align}
B(g,g') &= 2|\lambda| t
\intall \frac{d\Omega}{2\pi} \ln \frac{S(\Omega|g)+S(\Omega|g')}
{2\sqrt{S(\Omega|g)S(\Omega|g')}},
\end{align}
and find the Fisher information through the identity \cite{vantrees3}:
\begin{align}
\mathcal G(\omega_m) &= 4 \bk{\parti{g}{\omega_m}}^2\partit{}{g} B(g,g')\Big|_{g' = g}.
\end{align}
If the noise powers are quantum-limited,
\begin{align}
S_{\rm in} &= S' = 0.5,
\\
S(\Omega) &= V(\Omega) + 1.
\end{align}
After more algebra, we get
\begin{align}
\mathcal G(\omega_m) &= \frac{8|\lambda|^3t}{\omega_m^4} \intall \frac{d\Omega}{2\pi}\bk{\parti{V}{g}}^2
\frac{1}{(V+1)^2}.
\end{align}
Focusing on the OPO threshold, which occurs at
\begin{align}
g &= \bk{1-\Gamma^2/4}^{-1/2},
\end{align}
we obtain
\begin{align}
\mathcal G &= 
\frac{16\omega_m^2t}{\gamma^3}
\intall \frac{dx}{2\pi} \frac{1}{ (x^2+1)^2}
\frac{1}{[1+\Gamma^2x^2(x^2+1)]^2}.
\end{align}
To obtain an analytic result, suppose $\Gamma < 2$, such that we can
lower-bound $\mathcal G$:
\begin{align}
\mathcal G &> \frac{16\omega_m^2t}{\gamma^3}
\intall \frac{dx}{2\pi} \frac{1}{ (x^2+1)^2}
\frac{1}{[1+4x^2(x^2+1)]^2} 
\nonumber\\
&= \frac{1.532\omega_m^2t}{\gamma^3}.
\end{align}
In the limit of $\Gamma \to 0$, on the other hand,
$\mathcal G \to 4\omega_m^2 t/\gamma^3$, so
\begin{align}
\frac{4\omega_m^2 t}{\gamma^3}> \mathcal G >
 \frac{1.532\omega_m^2t}{\gamma^3},
\quad
0 <  \Gamma < 2,
\end{align}
which is the result quoted in the main text.  This result suggests
that the parameter estimation accuracy can be improved significantly
if $\gamma$ is reduced.

\begin{figure}[htbp]
\centerline{\includegraphics[width=0.45\textwidth]{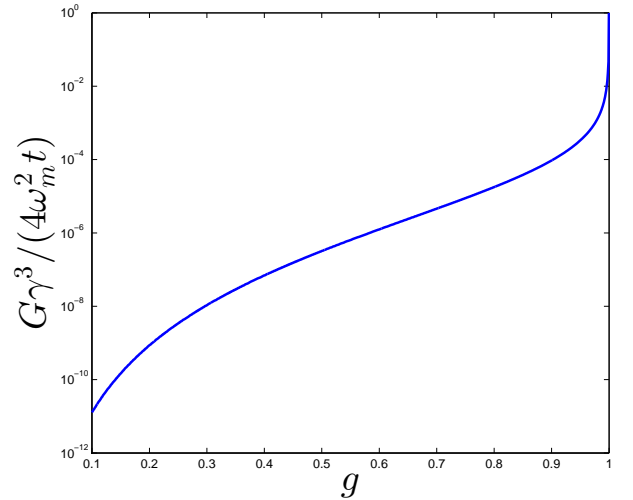}}
\caption{Normalized Fisher information versus $g$ on logarithmic scale
  for $\Gamma = 0.01$.}
\label{fisher}
\end{figure}

Below threshold, the Fisher information can be investigated by numerical integration
using this formula:
\begin{align}
\mathcal G = \frac{16\omega_m^2t}{\gamma^3\Gamma}
\intall \frac{d\Omega}{2\pi}\Bk{\Omega^2 -\bk{g^{-2}-1+\Gamma^2/4}}^2 \frac{V^4}{(V+1)^2}.
\end{align}
For example, Fig.~\ref{fisher} plots the normalized $\mathcal G$
versus $g$ on logarithmic scale for $\Gamma = 0.01$. The plot
demonstrates significant enhancement near $g = 1$.

So far all the results are derived for below-threshold operations. If
the perturbation causes the threshold to be exceeded, the system
becomes unstable, and we can no longer rely on the frequency-domain
analysis.  The analysis in the main text hints that instability should
improve the sensitivity even further, however.

\end{document}